\documentclass{article}
\usepackage{geometry}
\usepackage{amsmath,amssymb}
\usepackage[amsmath,thmmarks,framed]{ntheorem}
\usepackage{mathrsfs}
\usepackage{graphicx}
\usepackage{hyperref} 
\hypersetup{colorlinks,%
citecolor=black,%
filecolor=black,%
linkcolor=black,%
urlcolor=black}
\usepackage{setspace} 
\usepackage{latexsym}
\usepackage{graphicx}
\usepackage{tikz}
\usetikzlibrary{positioning}
\usetikzlibrary{shapes.geometric}
\usetikzlibrary{arrows,decorations.pathmorphing,backgrounds,positioning,fit,petri}                
\usepackage{framed}
\usepackage{setspace}
\usepackage{color}
\usepackage{cite}
\begin{document}

\begin{center}

 \begin{large}
 \textbf{A field theoretic model for static friction} \\
 \end{large}
 
 I. Mahyaeh and S. Rouhani\\
\textit{Department of Physics, Sharif University of Technology, Tehran, PO Box 11165-9161, Iran}\\
\end{center}

\textbf{Abstract}\\
We present a field theoretic model for friction, where the friction coefficient between two surfaces may be calculated based on elastic properties of the surfaces. We assume that the geometry of contact surface is not unusual. We verify Amonton's laws to hold that friction force is proportional to the normal load.This model gives the opportunity to calculate the static coefficient of friction for a few cases, and show that it is in agreement with observed values. Furthermore we show that the coefficient of static friction is independent of apparent surface area in first approximation.

\section{Introduction}
It is believed that friction between two solids comes from interlocking of their asperities \cite{Dowson,Rabinowicz} but this interlocking depends on the status of the two objects, whether they are at rest or have a relative motion. For the case of static friction, where two objects are at rest relative to each other with no lateral external force, there is an interlocking between two surfaces due to their roughness. In this state the load which is the combined gravitational force plus any vertical forces, deforms the object and the substrate at the microscopic scale. In order to cause an object to move, one should apply lateral force. Experiments have shown that motion starts once a critical lateral force is reached \cite{Dowson,Rabinowicz}.The amount of this critical force is proportional to load, the constant of proportionality being called the coefficient of static friction. Apparently this proportionality law goes back to Leonardo Da Vinci \cite{Dowson}. Later Amonton rediscovered the laws of friction; he understood friction as the force required to raise the surfaces pressing two bodies together. Later Belidor and Euler expanded on Amonton's work \cite{Dowson}. The understanding of friction was further developed by Coulomb who stressed four main factors: the nature of the materials in contact and their surface roughness; the extent of the surface area; the load; and the length of time that the surfaces remained in contact (time of repose) \cite{Dowson}. Coulomb also discusses the influence of sliding velocity, temperature and humidity. The distinction between static and dynamic friction is due to Segner \cite{Dowson}. The popular phenomenological laws of friction, which have been referred to as the Coulomb-Amonton laws state that: (i) Static Friction force is independent of the apparent area of contact. (ii) Static Friction is proportional to the normal load, with the coefficient of friction just depending on the nature of the two surfaces in contact.  (iii) Kinetic friction is same as above, however it does not depend on the sliding velocity but is smaller than static friction. Despite the universality of the friction laws, a satisfactory microscopic model for friction does not exist. Most microscopic models of friction are based on the concept of asperity, which refers to a protrusion of the surface \cite{Rabinowicz,Tabor}. So the two macroscopic surfaces are detached almost everywhere except at the tips of their asperities. Hence the true area of contact is given by the tips of asperities, which is much smaller than the apparent contact area.

Experimental evidence shows that Coulomb-Amonton laws are universal and the coefficient of friction depends on the nature of the interface alone\cite{Dowson}. However the universality of this law suggests that the static friction law should be derivable from a microscopic picture of the object and its interface with the substrate. 

A number of efforts in this direction have been made based on an assumption which expresses that friction is proportional to the real contact area, hence a relation between real contact area and load has been derived. The basic model is Hertz contact theory in which the contact area of a sphere($A$) subjected to an external load $L$ has been calculated within the theory of elasticity theory and shown to be proportional with $L^{2/3}$ \cite{Landau}. 

Archard \cite{Archard} proposes a model in which a sphere of radius $R_1$ has very small spheres on it's surface with radius $R_2 \ll R_1$, each sphere with radius $R_2$ has has very small spheres on it's surface with radius $R_3 \ll R_2$ and so on. This model is an attempt to incorporate different scales of roughness. If one considers just the first sphere, the well-known Hertz relation is derived, $A \sim L^{2/3}$; by cosidering scond spheres one comes to $A \sim L^{8/9}$. Finally adding the third spheres the relation changes to $A \sim L^{26/27}$. It seems that by considering infinite numbers of spheres which represent different scales of roughness on an object, one gets a direct proportionality between real contact area and load. 

Greenwood and Williamson \cite{Greenwood} investigated the contact between a plane and a nominally flat surface in which the tip of asperities assumed to be spherical with the same radii of curvature. They supposed that there is a probability distribution for the height of each asperity, asperities are disconnected mechanically and the contact between each asperity and the plane obeys Hertz relation. They considered exponential and Gaussian distributions; in the exponential case straight forward calculations show that real contact area is proportional to load but in the Gaussian distribution case they did numerical analysis and found the linear proportionality between real contact area and load for some cases.   

Muser et al \cite{Muser01} assumed exponential interactions between two surfaces in contact. Lateral and normal forces can be drived via differentiaing with respect to $x_{t}$ (the lateral posiotion of top surface) and $z_{t}$ (the mean height of top surface) respectively and the coefficient of static friction was defined as the maximum ratio of lateral force to the normal force. For the case of exponential potential it is is easy to see that coefficient of friction is independent of load but their numerical analysis shows the same result for Lennard-Jones potentials too. Also their model proposed that coefficiet of static friction for two commensurate surfaces is independent of contact area, reaches it's maximum for indentical surfaces and is independent of interaction strength. In another work \cite{Muser08} Muser constructed a field theory for contact between surfaces which was based on the assumption that the displacement field in the contact surface with substrate can be expressed as a function of surface profile. He calculated the probability distrbution of pressure and compared it with numerical analysis which show a good agreement.      

Persson \cite{Persson01,Perssonchem} considered the contact between two randomly rough surfaces and developed a theory for contacat in different length scales. Actually he calculated the real contact area with an arbitrary magnification and show that it is proportional to load for any rough surfaces provided that $\sigma_{0}=\dfrac{L}{A_{0}} \ll E^{*}$ where $L$ is load, $A_{0}$ is apparent contact area and $\dfrac{1}{E^{*}}=\dfrac{1-\nu_{1}^{2}}{E_{1}} + \dfrac{1-\nu_{1}^{2}}{E_{2}} $ where $E_{i}$ and $\nu_{i}$ are Young's modulus and Poisson ratio of materials respectively. The crucial quantity in Persson's theory is $G(\xi)$ which depends on $\sigma_{0}, E^{*}$ and surfaces's morphology and assumed to be much more than unit. He calculated  $G(\xi)$ for an special but important case of self-affine surfaces and one can see that it is possible have such surfaces which this quantity for them is comparable with unity. Hence, Persson's theory is restricted to cases for which $G(\xi) \gg 1$ and is not univeral.  
  
In this paper we present a fresh look at this problem based on the field theoretic method (assuming that scales are large enough so that quantum phenomena can be ignored). Consider an object on a rough substrate. Two surfaces are interlocked together. One should exert lateral force to bring the object out of interlocking, so the object can then move on the substrate. The lateral force deforms the object in the microscopic scale; hence it should be large enough to make an appropriate microscopic deformation. In this work we assume that substrate is rigid enough that it does not deform. This is obviously an invalid assumption but simplifies our calculations. We believe this assumption not make any fundamental change in the problem and we will discuss it in Section 4. We then go on to calculate the amount of work necessary to make this change, resulting in the critical lateral force. Physically this is equivalent to calculating the amount of change in the free energy of a system as result of a change in the boundary condition. As a simpler example one may consider an electrostatic problem, where a given boundary condition results in an electric field, leading to an energy density. Now attempting to change the boundary condition will face a resistance. In our interpretation this is precisely the friction force. 

This paper is organized as follows: In section 2 we present our model based on a field theoretic framework. We show that under quite general circumstances the Amonton's laws hold, at least in the first approximation. In section 3, we use our model to calculate a few coefficient of static friction and show that there is qualitative agreement with observed experimental values. In section 4, we discuss our results and assumptions. 

\section{The model}
For the case of static friction, consider an object at rest on a substrate with no lateral external force. Deformation of the interface is caused by the load and we assume that the substrate does not deform. The deformation depends on the geometry of the substrate and interactions between the object and the substrate. Exertion of a lateral force deforms the object further. We call the situation just after the exertion of lateral force state 1($t=0^+$), and when the object is just starting to move, state 2 ($t=T$). We expect that for transition from the first to the second state, the object makes microscopic slips on the substrate and undergoes some deformation in it’s structure and interface. To study these deformations consider the displacement element at a given point $u_\alpha (\textbf{x})$, Then the strain tensor is defined by,

\begin{equation}
u_{\alpha\beta}:=\dfrac{1}{2}(\dfrac{\partial u_\alpha}{\partial x_\beta} + \dfrac{\partial u_\beta}{\partial x_\alpha})
\end{equation}  
The Hamiltonian of an elastic isotropic material, up to second order in $u_{\alpha\beta}$ is  \cite{Landau,Kardar}

\begin{equation}
\label{elastichamiltonian}
H_{elasticity}= \dfrac{1}{2} \int (2\mu u_{\alpha\beta} u_{\alpha\beta} + \lambda u_{\alpha\alpha}u_{\beta\beta}) \mathrm{d}^3x 
\end{equation}

where $\mu$ and $\lambda$ are lame' coefficients. 

Let us now examine the boundary conditions. If we have some boundary forces, we can add another term to Hamiltonian, together with any bulk external interaction, like gravitational potential,

\begin{equation}
H_{bb}=\int B_\alpha (\textbf{x})u_\alpha (\textbf{x}) \mathrm{d}^3 x 
\end{equation}

Total haniltonain is just the summation of elastic and "boundary and bulk" Hamiltonian,

\begin{equation}
H = H_{elasticity} + H_{bb}
\end{equation}

Variation of the total Hamiltonian with respect to displacement field should be vanished, 

\begin{equation}
\dfrac{\delta H}{\delta u_\alpha (\textbf{x})} =0
\end{equation}
 
which yields the equation for equilibrium,

\begin{equation}
\label{equationofmotion}
\partial_\beta \sigma_{\alpha\beta}(\textbf{x})=B_\alpha (\textbf{x})
\end{equation}

in which $\sigma_{\alpha\beta}$ is the stress tensor,

\begin{equation}
\sigma_{\alpha\beta} = 2\mu u_{\alpha\beta} + \lambda u_{\gamma\gamma} \delta_{\alpha\beta}
\end{equation}

\begin{figure}
\centering
\includegraphics[scale=.8]{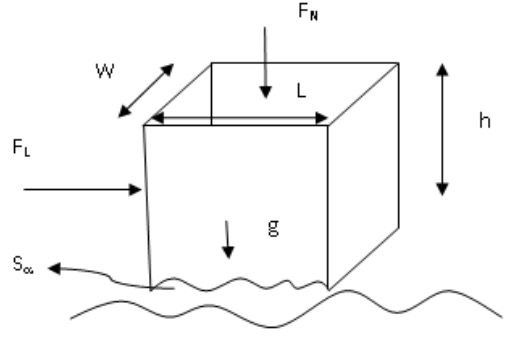}
\caption{Typical object with external forces}
\label{fig1}
\end{figure}

Since all the boundary forces are included in $B_\alpha (\textbf{x})$, see Fig.\ref{fig1}, solving Eqn.\eqref{equationofmotion} guarantees that forces are balanced in the boundary surfaces,

\begin{equation}
\label{boundary1}
\sigma_{\alpha\beta} (\textbf{x}) n_\beta (\textbf{x})=p_\alpha (\textbf{x})
\end{equation}
 
 in which $\textbf{x}$ is a point on the surface and $n_\beta (\textbf{x})$ is normal outward vector to the surface and $p_\alpha (\textbf{x})$ is the external pressure exerted on the surface.
 
   But one more boundary condition is needed; for the contact surface the displacement field $S_\alpha(x,y)$ of the object is induced by the substrate is given and fixed, so we have the boundary condition:
 
 \begin{equation}
 \label{boundary2}
 u_\alpha(x,y,0)=S_\alpha(x,y)
\end{equation}  

All temperature effects including expansion of the object are ignored to this order. Solving Equation of motion, Eqn.\eqref{equationofmotion},with given boundary conditions, i.e. Eqns.\eqref{boundary1} and \eqref{boundary2}, we get: 

\begin{equation}
\label{usolution}
u_{0,\alpha}(\textbf{q})=(2\pi)^3C_\alpha(\textbf{q}) - \int R_{\alpha\beta}(\textbf{q},\textbf{q}') B_\beta (-\textbf{q}')\mathrm{d}^3 q'
\end{equation}

in which
\begin{eqnarray}
u_\alpha(\textbf{q})=\int e^{-\imath \textbf{q}.\textbf{x}} u_\alpha(\textbf{x})\mathrm{d}^3x\\
B_\alpha(\textbf{q})=\int e^{-\imath \textbf{q}.\textbf{x}} B_\alpha(\textbf{x})\mathrm{d}^3x\\
C_\alpha(\textbf{q})=S_\gamma(\textbf{k})D_{\gamma\beta}(\textbf{k})f_{\beta\alpha}(\textbf{q})
\end{eqnarray}

and $ R_{\alpha\beta}(\textbf{q},\textbf{q}') $ defined as,

\begin{equation}
\label{greenfunction}
R_{\alpha\beta}(\textbf{q},\textbf{q}')=G_{\alpha\beta}(\textbf{q},\textbf{q}')-f_{\alpha\theta}(\textbf{q})D_{\theta\gamma}(\textbf{k})\delta(\textbf{k}+\textbf{k}')f_{\gamma\beta}(\textbf{q}')
\end{equation}

This leads to the Green function:
\begin{eqnarray}
f_{\alpha\beta}(\textbf{q}) = \dfrac{1}{\mu q^2} (\delta_{\alpha\beta}-\dfrac{\mu+\lambda}{2\mu+\lambda}\dfrac{q_\alpha q_\beta}{q^2})\\
G_{\alpha\beta}(\textbf{q},\textbf{q}')=f_{\alpha\beta}(\textbf{q})\delta(\textbf{q}+\textbf{q}')
\end{eqnarray}

\begin{equation}
D_{\alpha\beta}(\textbf{k})=\dfrac{1}{1- \dfrac{1}{2}\dfrac{\mu+\lambda}{2\mu+\lambda}} \dfrac{\mu k}{\pi} 
\left(
\begin{array}{ccc}
1- \dfrac{1}{2}\dfrac{\mu+\lambda}{2\mu+\lambda} \dfrac{k_y^2}{k^2} & \dfrac{1}{2}\dfrac{\mu+\lambda}{2\mu+\lambda} \dfrac{k_x k_y}{k^2} & 0\\
\dfrac{1}{2}\dfrac{\mu+\lambda}{2\mu+\lambda} \dfrac{k_x k_y}{k^2} & 1- \dfrac{1}{2}\dfrac{\mu+\lambda}{2\mu+\lambda} \dfrac{k_x^2}{k^2} & 0\\
0 & 0 & 1

\end{array} \right)
\end{equation}

The vector $\textbf{q}$ can be splited in to components along the surface $k_x , k_y$ and normal to the surface $q_z$. 
So;
\begin{equation}
\label{Sfouriertransform}
S_\alpha(\textbf{k})=\dfrac{1}{(2\pi)^2}\int e^{-\imath \textbf{k}.\textbf{r}} S_\alpha(\textbf{x})\mathrm{d}^2r
\end{equation}

The solution as offered by Eqn.\eqref{usolution} mainly has two parts; the one which is dependent on external forces $B_\alpha(\textbf{x})$ and have $R_{\alpha\beta}$ as Green function and one which is dependent on the displacement field $S_\alpha(x,y)$ of the object which is induced by the substrate. One can verify that $G_{\alpha\beta}$ is the Green function of the elasticity Hamiltonian, Eqn.\eqref{elastichamiltonian}, for the free space, so as we expect it has a contribution in the solution. If one imposes a boundary condition by which demands that displacement field on the contact surface should be zero, we need to add another Green function which comes from this boundary condition; this is (in abbreviation) $fDf$ term in Eqn.\eqref{greenfunction}. Now if one demands to change the last boundary condition and have an arbitrary displacement field on the contact surface with substrate, $S_\alpha(x,y)$, the term $(2\pi)^3C_\alpha(\textbf{q})$ should be added.
We insert this solution into the total Hamiltonian, and get: 
\begin{align}
\label{totalenergy}
H_0[B,S] &= -\dfrac{1}{2 (2\pi)^3} \int B_\alpha(-\textbf{q}) R_{\alpha\beta}(\textbf{q},\textbf{q}') B_\alpha(-\textbf{q}') \mathrm{d}^3q \mathrm{d}^3q'  \nonumber \\
& + \int C_\alpha(\textbf{q}) B_\alpha(-\textbf{q}) \mathrm{d}^3q \nonumber \\
& + \dfrac{1}{2} (2\pi)^3 \int S_\alpha(\textbf{k})D_{\alpha\beta}(\textbf{k})S_{\beta}(-\textbf{k}) \mathrm{d}^2k
\end{align}
 
The last term is usually used for elastic energy stored in an elastic block for the cases in which just the surface contact is considersed\cite{Persson06,Tosatti}. 
 One can verify by using  $u_0=\dfrac{\delta H_0[B,S]}{\delta B}$, that Eqn.\eqref{usolution} is yielded.
 
 As mentioned before, there is a transition from rest to the state which is appropriate for the object to move. In our model there is no change in $B_\alpha(\textbf{x})$ during the transition, but displacement field which is induced by the substrate changes, i.e.$S_\alpha(x,y)$. Intuitively we expect that there is a large deformation on the contact surface at rest which results from interlocking between the object and the substrate. When the object starts to move as a first step the interlocking is reduced, hence the deformation which is induced by substrate changes. The transition consists of two parts; the part in which the object slips on the substrate, hence the interlocking is reduced and the part in which the boundary condition on the contact surface changes. For slipping we consider the displacement vector as $l\widehat{x} + \omega\widehat{z}$, so any infinitesimal part of the object which is in $\textbf{x} + \textbf{u}_1(\textbf{x})$ at first state goes to $\textbf{x}+l\widehat{x} + \omega\widehat{z} + \textbf{u}_2(\textbf{x})$. From now on we forget the substrate, we just have an object with forces on it. We have some external forces included in $B_\alpha(\textbf{x})$ and there is pressure by something else which is $p_\alpha(\textbf{x})$, as given by Eqn.\eqref{boundary1}. Since the object is in equilibrium during the transition, the total work on the object due to slipping displacement is zero. The only remained part is the work which is done due to the change in the boundary condition, i.e. change in $\textbf{u}_2(\textbf{x})-\textbf{u}_1(\textbf{x})$. This displacement is not uniform throughout the object and yields nontrivial terms for work and change in internal energy. The change in this boundary condition changes the energy of the system which is obvious from Eqn.\eqref{totalenergy}. This change comes from the work which is done by $p_\alpha(\textbf{x})$ (i.e. work done by friction). We can formulate this by using the fact that
 
 \begin{equation}
 \label{externalforce}
 \dfrac{\delta H}{\delta S_\alpha}= \sigma_{\alpha\beta}n_\beta \Big \vert_{z=0}
\end{equation}   

And the RHS is just the pressure from the substrate. So we consider a change in boundary condition from $S_{1,\alpha}(x,y)$ to $S_{2,\alpha}(x,y)$ and using Eqns.\eqref{totalenergy} and \eqref{externalforce}, we get: 

\begin{equation}
\label{work}
H_0[B,S_2]-H_0[B,S_1]=\int_{S_1}^{S_2} \dfrac{\delta H}{\delta S_\alpha} \delta S_\alpha
\end{equation}

By defining $\Delta S_\alpha(x,y) = S_{2,\alpha}(x,y) - S_{1,\alpha}(x,y)$ and it’s Fourier transform as in Eqn.\eqref{Sfouriertransform},one can calculate the RHS of Eqn.\eqref{work}. This work is done by the substrate on the object. We assume that change in $S_\alpha(x,y)$ is small so that the calculations can be done up to the first order in $\Delta S_\alpha(x,y)$. By combining the result of this calculation and using Eqns.\eqref{totalenergy} and \eqref{work} we obtain

\begin{align}
\label{equ1}
\Delta (CB) &= \dfrac{\imath}{2\pi} (2\mu + \lambda) \int q_z R_{z\alpha}(\textbf{q},\textbf{q}') B_\alpha(-\textbf{q}') \Delta S_z(-\textbf{k})\mathrm{d}^3q  \mathrm{d}^3q' \nonumber \\
&+\dfrac{\imath}{2\pi} \mu \int q_z R_{y\alpha}(\textbf{q},\textbf{q}') B_\alpha(-\textbf{q}') \Delta S_y (-\textbf{k}) \mathrm{d}^3q  \mathrm{d}^3q' \nonumber \\
&+\dfrac{\imath}{2\pi} \mu  \int q_z R_{x\alpha}(\textbf{q},\textbf{q}') B_\alpha(-\textbf{q}') \Delta S_x(-\textbf{k})\mathrm{d}^3q  \mathrm{d}^3q' + O (\Delta S^2) 
\end{align}
 
 Finally, Eqn.\eqref{equ1} shows a linear relation between forces which are exerted on the object. We use $F_L$ for the total lateral force with the distribution $P _ {F_L} (\textbf{x})$ on the object and $F_z$ for the vertical force with the distribution $P _ {F_z} (\textbf{x})$ (this force can be weight which is distributed whole through the body of the object or any external normal force for example on the top side of the object as shown in Fig.1). Hence Eqn.\eqref{equ1} can be written as,
 \begin{equation}
 \label{linear relation}
 a F_L=b F_z
 \end{equation} 
 So this model predict a linear relation between the critical lateral force and vertical external forces, this is the well-known Amonton's law for friction.
 Coefficients \textit{a} and \textit{b} are:
 \begin{align}
 \label{acoefficient}
 a &=\dfrac{\imath}{2\pi} (2\mu + \lambda) \int q_z R_{zx}(\textbf{q},\textbf{q}') P_{F_L}(-\textbf{q}') \Delta S_z(-\textbf{k})\mathrm{d}^3q  \mathrm{d}^3q' \nonumber \\
&+\dfrac{\imath}{2\pi} \mu \int q_z R_{yx}(\textbf{q},\textbf{q}') P_{F_L}(-\textbf{q}') \Delta S_y(-\textbf{k})\mathrm{d}^3q  \mathrm{d}^3q' \nonumber \\
&+\dfrac{\imath}{2\pi} \mu  \int q_z R_{xx}(\textbf{q},\textbf{q}') P_{F_L}(-\textbf{q}') \Delta S_x(-\textbf{k})\mathrm{d}^3q  \mathrm{d}^3q' \nonumber \\
& - \int (\Delta C)_x (\textbf{q})  P_{F_L}(-\textbf{q}) \mathrm{d}^3q 
\end{align}    
and
 \begin{align}
 \label{bcoefficient}
 b&=\dfrac{\imath}{2\pi} (2\mu + \lambda) \int q_z R_{zz}(\textbf{q},\textbf{q}') P_{F_z}(-\textbf{q}') \Delta S_z(-\textbf{k})\mathrm{d}^3q  \mathrm{d}^3q' \nonumber \\
&+\dfrac{\imath}{2\pi} \mu \int q_z R_{yz}(\textbf{q},\textbf{q}') P_{F_z}(-\textbf{q}') \Delta S_y(-\textbf{k})\mathrm{d}^3q  \mathrm{d}^3q' \nonumber \\
&+\dfrac{\imath}{2\pi} \mu  \int q_z R_{xz}(\textbf{q},\textbf{q}') P_{F_z}(-\textbf{q}') \Delta S_x(-\textbf{k})\mathrm{d}^3q  \mathrm{d}^3q' \nonumber \\
& - \int (\Delta C)_z (\textbf{q})  P_{F_z}(-\textbf{q}) \mathrm{d}^3q 
\end{align} 
For simplicity we confined ourselves to a situation in which the dominant change in $S_\alpha(x,y)$ comes from $\Delta S_z(x,y)$ i.e. changes in the normal direction only. In this situation Eqns.\eqref{acoefficient} and \eqref{bcoefficient} can be calculated and we obtain:
\begin{align}
\label{thelast}
&F_L  \int P _ {F_L} (\textbf{x}) \Delta S_z(\textbf{r}')\dfrac{3 z^2 (x' - x)}{(\vert \textbf{r} - \textbf{r}' \vert ^2+z^2)^{\frac{5}{2}}} \mathrm{d}^3x \mathrm{d}^2r' = \nonumber \\
& F_z \int P_{F_z} (\textbf{x}) \Delta S_z(\textbf{r}') [\dfrac{3 z^3}{(\vert \textbf{r} - \textbf{r}' \vert ^2 +z^2)^{\frac{5}{2}}} +2(1-2\nu) \dfrac{z}{(\vert \textbf{r} - \textbf{r}' \vert ^2 +z^2)^{\frac{3}{2}}} ] \mathrm{d}^3x \mathrm{d}^2r'
\end{align}
In which $\nu$ is Poisson's ratio. Note that integrals should be evaluated on the real contact area in which the deformation and change in boundary condition takes place.  

\section{Calculation of coefficient of friction } 

Amonton's laws of friction state that the maximum of static friction is proportional to load, $L$, and the coefficient of friction is independent of apparent contact area. Consider a simple model for an object in which the object has $N$ uncorrelated asperity; this is a so-called single-asperity model. On average the force on the top side of each asperity is $\dfrac{L}{N}$. By using Eqn.\eqref{thelast} one can calculate the critical force which is needed to make an asperity to move and just by summing up all the forces for different ones we reach  the total critical force, namely maximum of static friction. Although asperities have a size distribution, we can consider an average size for all asperities as a first approximation. Also, since the asperity's size is so small compared to the apparent surface area we can assume that the pressure distribution is uniform on the top side of the asperity and we can neglect the weight of the asperity; so in this case $F_z$ is Load, $L$ and due to Eqn.\eqref{linear relation} critical friction force is proportional to Load.

With these assumptions integrals in Eqn.\eqref{thelast} can be evaluated on one asperity. If we assume an average size of $d \times d$ for asperities, the coefficient of friction which is calculated based on the Eqn.\eqref{thelast} would be independent of asperity$'$s size. So any correction due to dependency on the area will be on the real contact area and comes from size distribution of asperities. Hence, this model satisfies all the Amonton’s laws for static friction.

 The only remaining part is the calculation of coefficient of friction. For this purpose we should assume a form for  $\Delta S_z(x,y)$. As a simple case we consider a Cosinusoidal form with wavelength $\lambda$. We expect that for two typical surfaces asperities have the same order sizes, so the induced shape has a compareable wavelenght with the size of asperities. We further calculate a few coefficients for the $\lambda = d $ in which a typical material wants to silde on the same one (See Table.\ref{comparison}) and observe that they are not far from experimental values. 

\begin{table}
\caption{Comparison between prediction of model and observed values of $\mu$ \cite{Oberg}}
\centering
\begin{tabular}{c c c c}
\hline \hline
Material & $\nu$ & $\mu_{model}$ & $\mu_{exp.}$ \\
\hline
Glass  &  0.2 - 0.3  &  0.93 - 1.0 &  0.9 - 1.0\\
Copper   &  0.35  & 0.90 & 1.0 \\
Aluminium  & 0.33  & 0.91 & 1.05 - 1.35 \\
Iron & 0.2 - 0.3 & 0.9 - 1.0 & 1.0 \\
Graphite & 0.15 & 1.1 & 0.5 - 0.8 \\ 
\hline
\end{tabular}

\label{comparison}
\end{table}

\section{Discussion}

We have shown that if we consider a two state model for static friction and define the maximum of static friction as a force which is needed to make the transition from rest to the state which is ready to move, we can derive a linear relation between the lateral force and external load; also by considering asperity whose deformation makes the change of state, coefficient of friction is independent of apparent contact area. These toghether means that our model satisfies all Amonton's laws for static friction. 

We assumed that the substrate is rigid and has no deformation, but this is not necessary since we have taken all the forces into account, hence works done on the object. Thus we have already included the terms which come from work by substrate. However this essentially means that the material of the substrate is the same as the object. In other words our model is consistent with friction of an object sliding on a substrate of its own material. 

In the calculation of coefficient of friction we should have some assumptions for $\Delta S_z(x,y)$. The simplest case is the one by which we have the same materials as an object and substrate. So in this case the asperities have the same average size and it is acceptable to suppose that $\Delta S_z(x,y)$ has a wavelength equal to asperity's size(See Table. \ref{comparison}). However this is clearly a simplifying assumption and in future work we intend to make a better analysis of this point. 

The dependence of coefficient of friction on $\lambda$ needs an independent study. However to get an estimate for its effect, we did calculations for one order of magnitude range of $\lambda$ and observed that the coefficient of friction decreases with decrease of it , i.e. for  $\nu$= 0:3 if $\lambda$ decreases one order of magitude the coefficient of friction decreases by a factor of 2. Perhaps this may model the effect of lubrication. One can think that in the lubrication process the effective $\lambda$ decreases by filling larger $\lambda$ by a fluid, hence the coefficient of friction decreases.

In Table.\ref{comparison} one see that our predictions are better for metals but have deviations for graphite. Our theory is entirely elastic. Archard, Greenwood and Williamson and part of Persson's theories are all based on theory of elasticity. As Archard expresses \cite{Archard} considering plastic flow for first loading of an object is possible, but at the end in the steady state deformations will be elastic. Also Greenwood and Williamson \cite{Greenwood} believed that "the contact between flat surfaces can be determined either by plastic or by elastic conditions [...], while for very smooth ones, contact will be entirely elastic." It seems that we can expect that our model have good predicitions for smooth surfaces and conatct between them. So it is reasonable to have good results for metals which can be considered as smooth surfaces.

At the end we should emphasis that $\lambda$ is not the roughness wavelenght which has a range, but is the most important(or effective) wavelength in the change of the state.

\section{Acknowledgments}

We are indebted to Saman Moghimi for illuminating discussions and one of us I.M. to Daniel Bonn for hospitality in the University van Amsterdam whilst this work was in progress. We are also thankfull to Martin H. M{\"u}ser for reading an earlier version of this manuscript and very constructive comments.

\end{document}